\documentclass[a4paper,12pt]{article}
\usepackage{amsfonts,slashed}
\usepackage{url}
\usepackage{latexsym}
\usepackage{amsmath}
\usepackage{amssymb}
\usepackage{indentfirst}
\usepackage{graphicx}
\usepackage{epsfig}
\usepackage{amsthm}
\usepackage{amsfonts}
\usepackage{indentfirst}
\usepackage[colorlinks]{hyperref}
\usepackage{cite}
\usepackage{cancel}
\usepackage{tikz}
\usepackage{enumitem}

\begin{document}

\begin{titlepage}

%\date{}

\title{\bf{Supersymmetric near-horizon geometry and Einstein-Cartan-Weyl spaces}} 
\author{Dietmar Silke Klemm$^{1,2}$\thanks{dietmar.klemm@mi.infn.it} and Lucrezia Ravera$^{2}$\thanks{lucrezia.ravera@mi.infn.it} \\ \\
{\small $^{1}$\textit{Dipartimento di Fisica, Universit\`{a} di Milano, Via Celoria 16, 20133 Milano, Italy}}\\
{\small $^{2}$\textit{INFN, Sezione di Milano, Via Celoria 16, I-20133 Milano, Italy}}}
\clearpage\maketitle
\thispagestyle{empty}

\begin{abstract}
We show that the horizon geometry for supersymmetric black hole solutions of minimal five-dimensional 
gauged supergravity is that of a particular Einstein-Cartan-Weyl (ECW) structure in three dimensions, 
involving the trace and traceless part of both torsion and nonmetricity, and obeying some precise
constraints. In the limit of zero cosmological constant, the set of nonlinear partial differential equations 
characterizing this ECW structure reduces correctly to that of a hyper-CR Einstein-Weyl structure
in the Gauduchon gauge, which was shown by Dunajski, Gutowski and Sabra to be the horizon geometry
in the ungauged BPS case.
\end{abstract}

\vspace{2cm}

\noindent

\end{titlepage}

\section{Introduction}

More than forty years ago Hawking proved his famous theorem \cite{Hawking:1971vc,Hawking:1973uf}
on the topology of black holes, which asserts that event horizon cross sections of 4-dimensional
asymptotically flat stationary black holes obeying the dominant energy condition are topologically
$\text{S}^2$\footnote{In four dimensions, one can have black holes with nonspherical horizons by
relaxing some of the assumptions that go into Hawking's theorem. For instance, in asymptotically
anti-de~Sitter (aAdS) space, the horizon of a black hole can be a compact Riemann surface $\Sigma_g$
of any genus $g$ \cite{Lemos:1994xp}, or a sphere with two
punctures\cite{Gnecchi:2013mja,Klemm:2014rda}. In the latter case, the horizon is noncompact but
has yet finite area. For aAdS spaces, both the asymptotically flat and dominant energy
conditions are violated.}.
This result extends to outer apparent horizons in black hole spacetimes that are
not necessarily stationary \cite{Hawking:72}. Such restrictive uniqueness theorems do not hold in higher
dimensions, the most famous counterexample being the black ring of Emparan and Reall
\cite{Emparan:2001wn}, with horizon topology $\text{S}^2\times\text{S}^1$\footnote{Nevertheless,
Galloway and Schoen \cite{Galloway:2005mf} were able to show that, in arbitrary
dimension, cross sections of the event horizon (in the stationary case) and outer apparent horizons
(in the general case) are of positive Yamabe type, i.e., admit metrics of positive scalar curvature.}.

Moreover, in $d>4$ it is highly nontrivial to determine whether a given near-horizon geometry can be 
extended to a full black hole solution, since the strong uniqueness theorems
that hold in four dimensions \cite{Israel:1967wq,Carter:1971zc,Hawking:1971vc,Robinson:1975bv, Israel:1967za,Mazur:1982db} break down and there exist different black holes with the same asymptotic 
charges and different black hole solutions with the same near-horizon geometry.

In the last decade there has been significant progress in classifying near-horizon geometries, see
e.~g.~\cite{Kunduri:2006uh,Kunduri:2008rs,Gutowski:2008ca,Gutowski:2009wm,Grover:2013ima,
Gutowski:2013kma,Dunajski:2016rtx,Alaee:2018unn}.
In particular, the authors of \cite{Grover:2013ima} showed that for minimal gauged
five-dimensional supergravity the latter are at least half-supersymmetric. If they preserve a larger fraction
of supersymmetry, then they are locally isometric to AdS$_5$ with vanishing two-form field strength.
In the ungauged case of the same theory, it was recently shown in \cite{Dunajski:2016rtx} that
supersymmetric horizon geometries are given by three-dimensional Einstein-Weyl structures of
hyper-Cauchy-Riemann (hyper-CR) type \cite{Calderbank:1999ad,Dunajski:1999qs,Dunajski:2003zq, 
Dunajski:2013ee,Dunajski:2018qnr}\footnote{Notice that hyper-CR Einstein-Weyl spaces have originally
been called `special' in \cite{GauduchonTod}, and they have also been referred to as `Gauduchon-Tod 
spaces' in the literature, cf.~also \cite{Grover:2009ms}.}. In particular, it was proven in 
\cite{Dunajski:2016rtx} that a class of solutions of minimal supergravity in five dimensions is given by lifts
of three-dimensional Einstein-Weyl structures of hyper-CR type, and this class was characterized as the
most general near-horizon limit of supersymmetric solutions to the five-dimensional theory.
Moreover, it was deduced that a compact spatial section of a horizon can only be a Berger sphere, a
product metric on $\text{S}^1\times\text{S}^2$, or a flat three-torus. Subsequently, \cite{Dunajski:2016rtx}
considered the problem of reconstructing all supersymmetric solutions from a given near-horizon
geometry, and proved that the moduli space of infinitesimal supersymmetric transverse deformations of the near-horizon data is finite-dimensional if the spatial section of the horizon is compact. This analysis was 
carried on along the same lines of the one done in \cite{Li:2015wsa} for the case of nonsupersymmetric 
vacuum horizons in presence of a cosmological constant; see also \cite{Fontanella:2016lzo}.

In this paper, we extend some of the results of \cite{Dunajski:2016rtx} to minimal gauged supergravity
in five dimensions. In particular, we show that the horizon geometry of supersymmetric black holes
in this theory is that of a particular Einstein-Cartan-Weyl structure in three dimensions, which involves
the trace and traceless part of both torsion and nonmetricity, and obeys some precise constraints.
We also study the limit of zero cosmological constant, in which the set of nonlinear partial differential 
equations characterizing this ECW structure reduces correctly to that of a hyper-CR Einstein-Weyl
structure in the Gauduchon gauge, which was shown in \cite{Dunajski:2016rtx} to be the horizon
geometry in the ungauged BPS case. Moreover, it turns out that in the ungauged theory the geometry
of the horizon can be alternatively interpreted as a particular Einstein-Cartan-Weyl structure in three 
dimensions subject to some specific constraints.

The remainder of this paper is organized as follows: In section \ref{rev5dsugra} we briefly introduce
the theory of minimal gauged supergravity in five dimensions, and give a summary of the equations
satisfied by the near-horizon limit of supersymmetric black holes, following \cite{Grover:2013ima}.
In section \ref{ECW} we review the basic notions of Einstein-Cartan-Weyl geometry in three
dimensions \cite{Klemm:2018bil}. The main results are contained in section \ref{main}, where we
derive the correspondence between ECW structures in three dimensions and the near-horizon limit of 
supersymmetric black holes, and consider the limit $\Lambda\to 0$. We conclude in \ref{conclusions}
with some final remarks.

\section{$N=2$, $d=5$ gauged supergravity and the near-horizon limit of BPS black
holes}\label{rev5dsugra}

The bosonic action of minimal $N=2$, $d=5$ gauged supergravity is given 
by \cite{Grover:2013ima,Gunaydin:1984ak}\footnote{We use mostly plus signature.}
\begin{equation}\label{5dac}
S = \frac1{4\pi G}\int\left[\frac14\left(R + \frac{12}{\ell^2}\right)\star_5 1 - \frac12 F\wedge\star_5 F - 
\frac2{3\sqrt3} F\wedge F\wedge A\right],
\end{equation}
where $F=dA$ is a $\text{U}(1)$ field strength, $\ell$ is related to the cosmological constant by
$\Lambda=-6/\ell^2$, and $\star_5$ denotes the Hodge endomorphism in five dimensions. We adopt
the conventions of \cite{Grover:2013ima}. The equations of motion following from \eqref{5dac} read
\begin{equation}
R_{\alpha\beta} - 2 F_{\alpha\gamma}{F_\beta}^{\gamma} + \frac13 g_{\alpha\beta}\left(F^2 +
\frac{12}{\ell^2}\right) = 0\,, \qquad
d\star_5 F + \frac2{\sqrt3} F\wedge F =0\,, \label{eom-5d}
\end{equation}
with $F^2\equiv F_{\alpha\beta} F^{\alpha\beta}$.

\subsection{Near-horizon geometry of BPS black holes}\label{nhgeom}

To describe the near-horizon geometry it is convenient to introduce Gaussian null
coordinates $(u,r,y^i)$ \cite{Moncrief:1983xua, Friedrich:1998wq, Moncrief:2008mr}, defined in a 
neighborhood of a Killing horizon, where $g(V,V)=0$, with $V=\partial_u$ a Killing vector.
The horizon is then located at $r=0$, and $y^i$ are local coordinates on a
three-dimensional Riemannian manifold $\Sigma$ with metric $\gamma$, which is the spatial cross
section of the horizon. The metric and the two-form field strength are given by
\begin{align}
& ds^2 =  2 \mathbf{e}^+ \mathbf{e}^- + \gamma_{ij}dy^i dy^j\,, \nonumber \\
& F =  - \frac{\sqrt{3}}{2} \Phi \mathbf{e}^+ \wedge \mathbf{e}^- - \frac{\sqrt{3}}{2} r \mathbf{e}^+ \wedge \left(d\Phi - h \Phi\right) + dB\,,
\end{align}
with
\begin{equation}
\mathbf{e}^+ = du\,, \qquad\mathbf{e}^- = dr + rh - \frac12 r^2\Delta du\,,
\end{equation}
where the scalars $\Delta,\Phi$, the one-forms $h,B$, and the Riemannian metric $\gamma$
depend only on $y^i$ ($i,j=1,2,3$), cf.~\cite{Fontanella:2016lzo, Dunajski:2016rtx, Grover:2013ima} for more details. The one-form gauge potential associated to $F$ reads
\begin{equation}
A = \frac{\sqrt3}2 r\Phi du + B\,.
\end{equation}
The orientation is specified by
\begin{equation}
\epsilon_5 = \mathbf{e}^+ \wedge \mathbf{e}^- \wedge \epsilon_3\,,
\end{equation}
where $\epsilon_5$ is the five-dimensional volume form and $\epsilon_3$ is the volume form on $\Sigma$. 

In the near-horizon limit, the bosonic field equations \eqref{eom-5d} boil down to a set of equations
on the three-dimensional manifold $\Sigma$ \cite{Grover:2013ima}. In particular, from the gauge field equations one obtains 
\begin{equation}\label{gaugefield}
d \star_3 dB + \frac{\sqrt{3}}{2}\star_3\left(d\Phi - \Phi h\right) - h\wedge\star_3 dB - 2\Phi dB = 0\,,
\end{equation}
with $\star_3$ the Hodge dual on $\Sigma$. The nontrivial components of the Einstein
equations, namely $(ur)$ and $(ij)$, become respectively
\begin{align}
& \frac{1}{2} \nabla^i h_i  - \frac{1}{2}h^2  + \frac{1}{3}d B_{mn}dB^{mn}+ \Phi^2 - \Delta+ \frac{4}{\ell^2}=0\,, \label{ein+-} \\
& R_{ij} + \nabla_{(i}h_{j)} - \frac{1}{2}h_i h_j - 2 dB_{ik} d B_j^{\;k}+ \gamma_{ij} \left( \frac{1}{3}dB_{kl} dB^{kl} - \frac{1}{2}\Phi^2 + \frac{4}{\ell^2} \right) =0\,, \label{einij}
\end{align}
where $R_{ij}$ denotes the Ricci tensor on $\Sigma$, $h^2\equiv h_i h^i$, and $\nabla$ is the Levi-Civita 
connection of the metric $\gamma$.

One can prove (see \cite{Grover:2013ima}) that the necessary and sufficient conditions for a near-horizon 
geometry to be a supersymmetric solution of minimal five-dimensional gauged supergravity are given by
\begin{align}
& \Delta = \Phi^2\,, \label{susy1} \\
& \left(\frac{1}{2}h + \frac{1}{\sqrt{3}} \star_3 dB \right)^2 =\frac{1}{\ell^2} \label{susy2}\,.
\end{align}
Along the same lines of \cite{Grover:2013ima}, one can then introduce a one-form $Z$ such that
\begin{align}
& \frac{1}{2} h + \frac{1}{\sqrt{3}} \star_3 dB = \frac{1}{\ell} Z\,, \label{hbz} \\
& Z^2 \equiv Z^i Z_i =1\,. \label{Z^2=1}
\end{align}
Furthermore one must have \cite{Grover:2013ima}
\begin{equation}\label{nablaiZj}
\nabla_i Z_j = \left(-\frac3{\ell} + h^m Z_m\right)\gamma_{ij} + \frac3{\ell} Z_i Z_j - Z_i h_j - \frac12
\Phi (\star_3 Z)_{ij}\,,
\end{equation}
with $(\star_3 Z)_{ij} = \epsilon_{ijk} Z^k$.
Then, by taking the exterior derivative of \eqref{hbz} and making use of the gauge field equation \eqref{gaugefield}, one finds the condition
\begin{equation}\label{monopolegen}
\star_3 dh = d\Phi - 2\Phi h - 2\sqrt3\Phi\star_3 dB\,.
\end{equation}
As we will see below, using \eqref{susy2}, equ.~\eqref{monopolegen} can be rewritten as a generalized 
monopole equation \cite{Hitchin:1982gh, Jones:1985pla, Dunajski:2000rf, Dunajski:2006un, Klemm:2015mga}.

\section{Einstein-Cartan-Weyl geometry in three dimensions}\label{ECW}

In this section, we briefly review Einstein-Cartan-Weyl geometry in three dimensions,
following \cite{Klemm:2018bil}. 

Consider a three-dimensional Einstein manifold endowed with a metric $\gamma$. The connection
$\hat\Gamma$, which is assumed to have nonvanishing torsion and nonmetricity, can be decomposed as
\begin{equation}\label{connection}
\hat{\Gamma}^l_{\; ij} = {\Gamma}^l_{\; ij} + N^l_{\; ij}\,,
\end{equation}
where $\Gamma$ denotes the Levi-Civita connection and $N^l_{\; ij}$ are the components of the distortion. 
The latter can be written as 
\begin{equation}\label{distortion}
N_{lij} = \frac12\left(T_{jli} - T_{lji} - T_{ijl}\right) + \frac12\left(Q_{lij} + Q_{lji} - Q_{ilj}\right)\,.
\end{equation}
Here $T^l_{\; ij}$ is the torsion, antisymmetric in the last two indices,
\begin{equation}\label{torsion}
T^l_{\; ij} = \hat\Gamma^l_{\;ij} - \hat\Gamma^l_{\; ji} ,
\end{equation}
while $Q_{lij}$ is the nonmetricity tensor,
\begin{equation}\label{nonmetricityECW}
Q_{lij} = -\hat{\nabla}_i\gamma_{jl}\,,
\end{equation}
where $\hat{\nabla}$ is the covariant derivative associated to $\hat\Gamma$. $Q_{lij}$ can be
decomposed into a trace and traceless part,
\begin{equation}\label{nonmetricityECWfull}
Q_{lij} = - 2\Theta_i\gamma_{jl} + \tilde{Q}_{lij}\,,
\end{equation}
with $\Theta_i$ the Weyl vector and ${{\tilde Q}^j}_{\,ij}=0$. In three dimensions, the decomposition
for the torsion reads
\begin{equation}\label{dector}
T^l_{\; ij} = \tilde{T}^l _{\; ij} + \frac12\left(\delta^l_{\; j} T_ i - \delta^l_{\; i} T_j \right) ,
\end{equation}
where $\tilde{T}^j_{\; ij }=0$ and $T_i \equiv T^j_{\; i j}$.
For simplicity of notation, let us define the traceless part of the distortion as
\begin{equation}\label{tracelessN}
\tilde{N}_{lij}\equiv\frac12\left(\tilde{T}_{jli} - \tilde{T}_{lji} - \tilde{T}_{ijl}\right) + \frac12\left(\tilde{Q}_{lij}
+ \tilde{Q}_{lji} - \tilde{Q}_{ilj}\right)\,,
\end{equation}
such that
\begin{equation}\label{distcontECWfull}
N_{lij} = \tilde{N}_{lij} + \Theta_l \gamma_{ij} - \Theta_i \gamma_{l j} - \Theta_j \gamma_{li}
+ \frac12\left(\gamma_{ij} T_l - \gamma_{il} T_j\right)\,.
\end{equation}
An Einstein-Cartan-Weyl space is defined as one for which the symmetrized Ricci tensor $\hat{R}_{(ij)}$
of $\hat{\nabla}$ is proportional to the metric. In particular, in three dimensions one has
\begin{equation}\label{cond}
\hat{R}_{(ij)} = \frac{1}{3} \hat{R} \gamma_{ij} ,
\end{equation}
where $\hat{R}$ denotes the scalar curvature of $\hat{\nabla}$. Under a Weyl rescaling
$\gamma_{ij}\mapsto e^{2\omega}\gamma_{ij}$, the one-form $\Theta$ and the connection
$\hat\Gamma$ transform according to
\begin{equation}\label{transfECW}
\Theta_i\mapsto\Theta_i + \xi\partial_i\omega\,, \qquad
\hat{\Gamma}^i_{\;jk}\mapsto\hat{\Gamma}^i_{\;jk} + (1-\xi)\delta^i_{\;k}\partial_j\omega\,,
\end{equation}
where $\xi$ denotes an arbitrary parameter that we are free to include \cite{Smalley:1986tr,Moon:2009zq}.
This means that the torsion and the nonmetricity tensor transform respectively as
\begin{equation}\label{transftorsandnonmetECW}
T^i_{\;jk}\mapsto T^i_{\;jk} + 2(1-\xi)\delta^i_{\;[k}\partial_{j]}\omega\,, \qquad
Q^i_{\;jk}\mapsto Q^i_{\;jk} - 2\xi\delta^i_{\;k}\partial_j\omega\,,
\end{equation}
which implies
\begin{equation}\label{transfTtildeTThetaECW}
T_i\mapsto T_i + 2(1-\xi)\partial_i\omega\,, \qquad\tilde{T}^i_{\;jk}\mapsto\tilde{T}^i_{\;jk}\,.
\end{equation}
For the Riemann tensor, the Ricci tensor and the scalar curvature one obtains
\begin{equation}\label{transfRECW}
\hat{R}^i_{\;jkl}\mapsto\hat{R}^i_{\;jkl}\,, \qquad \hat{R}_{ij}\mapsto\hat{R}_{ij}\,, \qquad
\hat{R}\mapsto e^{-2\omega}\hat{R}\,,
\end{equation}
and thus the condition \eqref{cond} is Weyl invariant.
In terms of Riemannian data, \eqref{cond} becomes \cite{Klemm:2018bil}
\begin{equation}\label{PDEECW}
\begin{split}
& R_{ij}  + \nabla_{(i} \Theta_{j)} + \Theta_i \Theta_j +\frac{1}{2} \nabla_{(i} T_{j)}+ \frac{1}{4}T_i T_j + \Theta_{(i} T_{j)} \\
& - \tilde{N}^{lm}_{\;\;\;(i} \tilde{N}_{j) lm} + \Theta^l \tilde{N}_{(ij)l}+ \frac{1}{2} T^l \tilde{N}_{(ij) l} - \nabla_l \tilde{N}_{(i \;\; j)}^{\;\;\;l}  \\
& = \frac{1}{3} \gamma_{ij} \left( R + \nabla^k \Theta_k + \Theta^k \Theta_k + \frac{1}{2} \nabla^k T_k + \frac{1}{4} T^k T_k + \Theta^k T_k - \tilde{N}^{lmn} \tilde{N}_{mnl} \right)\,,
\end{split}
\end{equation}
where $R_{ij}$ and $R$ are the Ricci tensor and the scalar curvature of the Levi-Civita connection.
The Ricci scalar for a three-dimensional Einstein-Cartan-Weyl manifold reads
\begin{equation}\label{rsECW}
\hat{R} = R + 4 \nabla^k\Theta_k - 2 \Theta^k \Theta_k  + 2 \nabla^k T_k  - \frac{1}{2} T^k T_k -2 \Theta^k T_k - \tilde{N}^{lmn} \tilde{N}_{mnl}\,.
\end{equation}
Finally, notice that we can define a one-form
\begin{equation}\label{thetacheck}
\check{\Theta}_i\equiv\Theta_i + \frac12 T_i\,,
\end{equation}
such that \eqref{PDEECW} and \eqref{rsECW} can be recast in the form
\begin{equation}\label{PDEECWnewNM}
\begin{split}
& R_{ij}  + \nabla_{(i} \check{\Theta}_{j)} + \check{\Theta}_i \check{\Theta}_j  - \tilde{N}^{lm}_{\;\;\;(i} \tilde{N}_{j) lm} + \check{\Theta}^l \tilde{N}_{(ij)l} - \nabla_l \tilde{N}_{(i \;\; j)}^{\;\;\;l} \\
& = \frac{1}{3} \gamma_{ij} \left( R + \nabla^k \check{\Theta}_k + \check{\Theta}^k \check{\Theta}_k  - \tilde{N}^{lmn} \tilde{N}_{mnl} \right)\,,
\end{split}
\end{equation}
\begin{equation}\label{rsECWnewNM}
\hat{R} = R + 4 \nabla^k \check{\Theta}_k - 2 \check{\Theta}^k \check{\Theta}_k - \tilde{N}^{lmn} \tilde{N}_{mnl}\,.
\end{equation}
We see that the traces $\Theta_i$ and $T_i$ appear only through the linear combination \eqref{thetacheck},
which transforms as $\check\Theta\mapsto\check\Theta+d\omega$ under \eqref{transfECW}. In fact,
a torsion trace can always be shuffled into a Weyl vector and vice versa, as can be easily seen from
the first Cartan structure equation\footnote{Such a reshuffling changes of course the definition
of parallel transport.}. This is the reason for the freedom to include the arbitrary parameter $\xi$
in \eqref{transfECW}.

\section{Three-dimensional ECW structures and $N=2$, $d=5$ gauged supergravity}\label{main}

In this section, we shew that the horizon geometry for supersymmetric black hole solutions of minimal
five-dimensional gauged supergravity is that of a particular Einstein-Cartan-Weyl structure in three 
dimensions. To this aim, consider the field equations \eqref{gaugefield}, \eqref{ein+-} and \eqref{einij}
and assume that the supersymmetry constraints \eqref{susy1} and \eqref{susy2} hold.
Thus, we have
\begin{equation}
dB_{jk} = - \frac{\sqrt{3}}{2} \epsilon_{ijk} h_i + \frac{\sqrt{3}}{\ell} \epsilon_{ijk} Z_i\,,
\end{equation}
and
\begin{align}
& dB_{im}dB_{j}^{\; m} = \frac34(\gamma_{ij} h^2 - h_i h_j) + \frac3{\ell^2} (\gamma_{ij} Z^2 - Z_iZ_j) -
\frac3{\ell} (\gamma_{ij} h^m Z_m - h_{(i} Z_{j)} )\,, \label{b1} \\
& dB_{im} dB^{im} = \frac32 h^2 + \frac6{\ell^2} Z^2 - \frac6{\ell}h^iZ_i\,. \label{b2}
\end{align}
Furthermore, using \eqref{susy2}, equ.~\eqref{monopolegen} can be recast in the form
\begin{equation}\label{genmon}
\star_3 \left[d\Phi + \left( h - \frac{6}{\ell}Z \right) \Phi \right] =dh\,,
\end{equation}
which is the generalized monopole equation \cite{Hitchin:1982gh, Jones:1985pla, Dunajski:2000rf, Dunajski:2006un}. Thus, the gauge field equation \eqref{gaugefield} reduces to the generalized monopole 
equation \eqref{genmon}. Note that $\Phi$ is a weighted scalar with conformal weight $-1$ on $\Sigma$.

The $(ur)$ component \eqref{ein+-} of the Einstein equations becomes
\begin{equation}\label{gengauduchon}
\nabla^i h_i = - \frac{12}{\ell^2}Z^i Z_i + \frac4{\ell} h^i Z_i\,.
\end{equation}
The symmetrized part and the trace part of \eqref{nablaiZj} give respectively
\begin{equation}\label{nzijsymmetric}
\nabla_{(i} Z_{j)} = \left(-\frac3{\ell} Z^m Z_m + h^m Z_m\right)\delta_{ij} + \frac3{\ell} Z_i Z_j -
Z_{(i} h_{j)}\,,
\end{equation}
\begin{equation}\label{nzii}
\nabla^i Z_i  = 2 h^i Z_i - \frac6{\ell} Z^i Z_i\,,
\end{equation}
and thus \eqref{gengauduchon} can be written as
\begin{equation}\label{nhnZ}
\nabla^i h_i =\frac{2}{\ell}\nabla^i Z_i\,,
\end{equation}
which generalizes the Gauduchon gauge $\nabla^i h_i=0$, that holds in the case of vanishing
cosmological constant, i.e., $\ell\to\infty$ \cite{Dunajski:2016rtx}.

Moreover, using \eqref{b1}, \eqref{b2} and \eqref{gengauduchon}, the $(ij)$-components \eqref{einij}
of the Einstein equations yield
\begin{equation}\label{ijcompsymm}
R_{ij} + \nabla_{(i} h_{j)} + h_i h_j + \frac6{\ell^2} Z_i Z_j - \frac6{\ell} h_{(i} Z_{j)} = \left(\frac12\Phi^2
+ h^k h_k - \frac4{\ell} h^k Z_k\right)\gamma_{ij}\,,
\end{equation}
whose trace, together with \eqref{gengauduchon}, leads to
\begin{equation}\label{R3d}
R = \frac12\left(3\Phi^2 + 4 h^i h_i + \frac{12}{\ell^2} Z^i Z_i - \frac{20}{\ell} h^i Z_i\right)\,.
\end{equation}
Observe that the limit $\ell\rightarrow\infty$ of \eqref{ijcompsymm} and \eqref{R3d} exactly reproduces
the results of \cite{Dunajski:2016rtx}, i.e., the conditions on the horizon geometry in the ungauged case, 
without cosmological constant.

To finally show that the horizon geometry for BPS black holes in minimal $d=5$ gauged 
supergravity is that of a particular Einstein-Cartan-Weyl structure in three dimensions,
consider a three-dimensional ECW space for which the following conditions hold:
\begin{enumerate}[label=\Alph*)]
\item There exists a scalar $\Phi$ of conformal weight $-1$ that, together with the nonmetricity and
torsion traces $\Theta$ and $T$, satisfies the generalized monopole equation
\begin{equation}\label{monECW1}
\star_3 \left(d\Phi + \check{\Theta}\Phi\right) = d\Theta\,.
\end{equation}
Notice that \eqref{monECW1} is invariant under
\begin{equation}
\Phi\mapsto e^{-\omega}\Phi\,, \qquad\Theta\mapsto\Theta + \xi d\omega\,, \qquad\check{\Theta}
\mapsto\check{\Theta} + d\omega
\end{equation}
for any value of $\xi$.
\item The trace part of the torsion satisfies
\begin{equation}\label{T^2}
T^2 \equiv T^i T_i = c^2\,,
\end{equation}
where $c$ is a constant, and
\begin{equation}\label{nablaiTj}
\nabla_i T_j = \left(\frac14 T^k T_k + \Theta^k T_k\right)\gamma_{ij} - T_i\Theta_j - \frac14 T_i T_j - 
\frac12\Phi\epsilon_{ijk} T^k\,,
\end{equation}
which implies in particular
\begin{equation}\label{nablaiTjsymm}
\nabla_{(i} T_{j)} = \left(\frac14 T^k T_k + \Theta^k T_k\right)\gamma_{ij} - T_{(i}\Theta_{j)} - \frac14
T_i T_j\,,
\end{equation}
and
\begin{equation}\label{nablaT}
\nabla^i T_i = 2\Theta^i T_i + \frac12 T^i T_i\,.
\end{equation}
\item The Weyl vector obeys
\begin{equation}\label{nablaTheta}
\nabla^i\Theta_i = - \frac13\Theta^i T_i - \frac1{12} T^i T_i\,.
\end{equation}
\item The traceless part of the torsion is totally antisymmetric and reads
\begin{equation}\label{TtildeECWdef}
\tilde{T}_{lmn} = \Phi\epsilon_{lmn}\,,
\end{equation}
while the traceless part of the nonmetricity is given by
\begin{equation}\label{QtildeECWdef}
\tilde{Q}_{mln}  = \frac{2c}{\sqrt3}\epsilon_{lk(m} T^k T_{n)}\,,
\end{equation}
and thus
\begin{equation}
\tilde N_{lmn} = \frac c{\sqrt3}\left(\epsilon_{lmk} T^k T_n + \epsilon_{lnk} T^k T_m\right) + \frac12
\Phi\epsilon_{lmn}\,. \label{NtildeECWdef}
\end{equation}
\item The Ricci scalar of the affine connection is
\begin{equation}\label{RhatECWdef}
\hat{R} = -\frac{3c}2\Theta^i T_i + \frac92 c^2\,.
\end{equation}
\end{enumerate}
Observe that in terms of $\hat\nabla$, \eqref{nablaiTj} reads
\begin{equation}
\hat\nabla_i T_j = \frac14 T_i T_j - \Phi\epsilon_{ijk} T^k - \frac14\gamma_{ij} T^k T_k + \Theta_i T_j\,,
\end{equation}
and thus
\begin{equation}
\hat\nabla_T T_j = \Theta^k T_k T_j\,,
\end{equation}
which means that the vector $u_j\equiv f^{-1} T_j$, where the function $f$ satisfies
$T^i\partial_i\ln f=\Theta^i T_i$, is parallel transported along its integral curves, $\hat\nabla_u u=0$.
Notice also that we can define a torsionful but metric connection $\bar\nabla$, with torsion
trace and traceless part respectively given by $\bar T_i=\frac12 T_i + 2\Theta_i$ and
$\tilde{\bar T}_{ijk}=-\Phi\epsilon_{ijk}$, such that \eqref{nablaiTj} becomes
\begin{equation}
\bar\nabla_i T_j = 0\,.
\end{equation}
This implies $\bar\nabla_{(i} T_{j)} = 0$, and therefore $T_j$ is a Killing vector with
torsion \cite{Kubiznak:2009qi}.

Note that the `gauge fixing' conditions \eqref{nablaT} and \eqref{nablaTheta} lead to
\begin{equation}\label{gengaudECW}
\nabla^i\Theta_i = -\frac16\nabla^i T_i\,.
\end{equation}
We now identify
\begin{equation}
\Theta = h\,, \qquad T = - \frac{12}{\ell} Z\,, \qquad c = \frac{12}\ell\,, \label{identif}
\end{equation}
such that \eqref{T^2} turns into \eqref{Z^2=1}, while \eqref{QtildeECWdef}, \eqref{NtildeECWdef} and 
\eqref{RhatECWdef} assume the form
\begin{align}
& \tilde{Q}_{mln}  = \frac{4\sqrt3}{\ell}\epsilon_{lk(m} Z^k Z_{n)}\,, \\
& \tilde{N}_{lmn} = \frac{2\sqrt3}{\ell} \left(\epsilon_{lmk} Z^k Z_n + \epsilon_{lnk} Z^k Z_m\right) +
\frac12\Phi\epsilon_{lmn}\,, \label{Ntildelast} \\
&\hat{R} = -\frac{18}{\ell} h^i Z_i + \frac{54}{\ell^2}\,. \label{Rhatlast1}
\end{align}
Moreover, under the identifications \eqref{identif}, eqns.~\eqref{nablaiTj}, \eqref{nablaiTjsymm}, 
\eqref{nablaT}, \eqref{nablaTheta} and \eqref{gengaudECW} become respectively \eqref{nablaiZj}, 
\eqref{nzijsymmetric}, \eqref{nzii}, \eqref{gengauduchon} and \eqref{nhnZ}.
Likewise, the generalized monopole equations \eqref{genmon} and \eqref{monECW1} coincide.

For the case we are considering, the Einstein-Cartan-Weyl equations \eqref{PDEECW} read
\begin{equation}\label{PDEECWour}
R_{ij}  + \nabla_{(i} h_{j)} + h_i h_j + \frac6{\ell^2} Z_i Z_j - \frac6{\ell} h_{(i} Z_{j)}= \frac13\gamma_{ij}
\left(R + h^k h_k - \frac6{\ell^2} Z^k Z_k - \frac2{\ell} h^k Z_k\right)\,.
\end{equation}
One can now use the expression \eqref{rsECW} for the Ricci scalar of the affine connection to rephrase
the constraint \eqref{R3d} as \eqref{Rhatlast1}. 
Finally, using \eqref{R3d}, we see that \eqref{ijcompsymm} is equivalent to the ECW equations
\eqref{PDEECWour}, that is the set of partial differential equations characterizing an Einstein-Cartan-Weyl
manifold in the gauge \eqref{nhnZ}, subject to the conditions \eqref{genmon} and \eqref{Rhatlast1},
together with the constraints on the trace part of the torsion (cf.~\eqref{T^2} and \eqref{nablaiTj}) and
on the traceless part of the distortion (cf.~\eqref{Ntildelast}). 

We have thus shown that the horizon geometry for supersymmetric black hole solutions of minimal
five-dimensional gauged supergravity is that of a particular Einstein-Cartan-Weyl structure in three
dimensions, in the gauge \eqref{nhnZ}, subject to the constraints \eqref{genmon}, \eqref{T^2}, 
\eqref{nablaiTj}, \eqref{Ntildelast} and \eqref{Rhatlast1}.
Notice that the conditions B) - E) above break conformal invariance, but this was to be expected, since
the supergravity theory we started with is not conformally invariant.

\subsection{Observations on the limit $\ell\rightarrow\infty$}

Let us finally comment on the special case when the cosmological constant goes to zero. As we have
already mentioned, the limit $\ell\rightarrow\infty$ of \eqref{ijcompsymm} and \eqref{R3d} exactly 
reproduces the results of \cite{Dunajski:2016rtx}, i.e.,
\begin{align}
& R_{ij} + \nabla_{(i} h_{j)} + h_i h_j = \left(\frac12\Phi^2 + h^k h_k\right)\gamma_{ij}\,, \label{ijcompsymmLINF} \\
& R = \frac12\left(3\Phi^2 + 4 h^k h_k\right)\,. \label{R3dLINF}
\end{align}
The same holds for the generalized monopole equation \eqref{genmon}, which, for $\ell\rightarrow\infty$, 
reduces to the monopole equation found in \cite{Dunajski:2016rtx}, that is
\begin{equation}\label{genmonLINF}
\star_3\left( d\Phi + h\Phi\right) = dh\,.
\end{equation}
Moreover, for $\ell\rightarrow\infty$, the conditions on the Einstein-Cartan-Weyl geometry of section 
\ref{main} boil down to
\begin{align}
& T_i =0\,, \label{TiLINF} \\
& \nabla^i h_i =0\,, \label{nablahLINF} \\
& \tilde{N}_{lmn} = \frac12\Phi\epsilon_{lmn}\,, \label{NtildelastLINF} \\
& \hat{R}= 0\,. \label{hatRLINF}
\end{align}
In particular, \eqref{nablahLINF} is called the Gauduchon gauge.
We can thus conclude that the horizon geometry for supersymmetric black holes in $d=5$
ungauged supergravity not only corresponds to a three-dimensional hyper-CR Einstein–Weyl structure in
the Gauduchon gauge (as shown in \cite{Dunajski:2016rtx}), but also to an Einstein-Cartan-Weyl structure
in the Gauduchon gauge and subject to the constraints \eqref{TiLINF} (vanishing torsion trace),
\eqref{genmonLINF}, \eqref{NtildelastLINF} (which defines the traceless part of the torsion, which is 
completely antisymmetric, while the traceless part of the nonmetricity is zero), and \eqref{hatRLINF}
(vanishing Ricci scalar of the affine connection). This ambiguity comes from the fact that the sets of
nonlinear partial differential equations characterizing the hyper-CR Einstein-Weyl structure
of \cite{Dunajski:2016rtx} and the ECW structure defined by \eqref{genmonLINF}-\eqref{hatRLINF} coincide.

\section{Conclusions}\label{conclusions}

It was shown recently in \cite{Dunajski:2016rtx} that the horizon geometry of BPS black holes in minimal 
$d=5$ supergravity defines a hyper-CR Einstein-Weyl structure.
Here, we extended this result to $N=2$, $d=5$ gauged supergravity, and showed that in this case
supersymmetric black hole horizons correspond to a particular three-dimensional Einstein-Cartan-Weyl 
structure, given in the gauge \eqref{nhnZ} and obeying the constraints \eqref{genmon}, \eqref{T^2}, 
\eqref{nablaiTj}, \eqref{Ntildelast} and \eqref{Rhatlast1}.

As a byproduct, it turned out that in the limit of vanishing cosmological constant, the horizon
geometry can be alternatively interpreted as an ECW structure subject to
\eqref{genmonLINF}-\eqref{hatRLINF}.

Future developments of our work include possible extensions to higher dimensions and to the
matter-coupled case.

\section*{Acknowledgements}

D.~S.~K. is supported partly by INFN.

\end{document}